\documentclass[a4paper,fleqn,usenatbib]{mnras}

\usepackage[T1]{fontenc}
\usepackage{ae,aecompl}

\usepackage{graphicx}	% Including figure files
\usepackage{amsmath}	% Advanced maths commands
\usepackage{amssymb}	% Extra maths symbols

\title{Clues to the nature of ultra diffuse galaxies from estimated galaxy velocity dispersions}

\author[D. Zaritsky]{
Dennis Zaritsky$^{1}$\thanks{E-mail: dennis.zaritsky@gmail.com}
\\
$^{1}$Steward Observatory, University of Arizona, Tucson, AZ 85719, USA
}

\date{Accepted 2016 September 23; Received 2016 September 03; in original form 2016 August 14}

\pubyear{2016}

\begin{document}
\label{firstpage}
\pagerange{\pageref{firstpage}--\pageref{lastpage}}
\maketitle

\begin{abstract}
We describe how to estimate the velocity dispersions of ultra diffuse galaxies, UDGs, using a previously defined galaxy scaling relationship. 
The method is accurate for the two UDGs with spectroscopically measured dispersions, as well as for ultra compact galaxies, ultra faint galaxies, and stellar systems with little or no dark matter. This universality means that the relationship can be applied without further knowledge or prejudice regarding the structure of a galaxy. We then estimate the velocity dispersions of UDGs drawn from two published samples and examine the distribution of total masses. We find, in agreement with the previous studies of two individual UDGs, that these systems are dark matter dominated systems, and that they span a range of at least $10^{10} < M_{200}/M_\odot < 10^{12} $. These galaxies are not, as an entire class, either all dwarfs or all failed $L_*$ galaxies. Estimates of the velocity dispersions
can also help identify interesting subsets of UDGs, such as those that are likely to have the largest mass-to-light ratios, for subsequent spectroscopic study.
\end{abstract}

\begin{keywords}
galaxies:  evolution -- kinematics and dynamics -- fundamental parameters
\end{keywords}

%%%%%%%%%%%%%%%%%%%%%%%%%%%%%%%%%%%%%%%%%%%%%%%%%%

%%%%%%%%%%%%%%%%% BODY OF PAPER %%%%%%%%%%%%%%%%%%

\section{Introduction}

Deep imaging surveys are uncovering extensive samples of low surface brightness galaxies. Some low luminosity examples are satellites of our own Galaxy \citep{bechtol,drlica} and others, much larger and more luminous, lie well beyond the Local Group \citep{vdk,mihos,koda,munoz,roman}. These objects potentially hold key clues on open questions as diverse as the nature of dark matter \citep{ackerman} and the drivers of galaxy formation \citep{agretz}. To understand these galaxies and utilise them in addressing these broader questions, we must measure
their masses. Are the distant low surface galaxies, broadly referred to as ultra diffuse galaxies or UDGs,  ``failed" massive galaxies or spatially extended dwarf galaxies \citep{vdk,beaslyb,amorisco}? 

A recent study required 33.5 hrs of exposure time on a 10m telescope to obtain the integrated light spectrum from which the line of sight velocity dispersion of
a single UDG was measured \citep{vdk16}. We are reaching the limit of what we can accomplish with current capabilities. Compiling large samples of such galaxies with measured internal kinematics is beyond what we can hope to do. How to choose the systems on which to spend our valuable resources most fruitfully?

We propose exploiting galaxy scaling relations that relate photometric and kinematics parameters. Such relations also depend on the distance to the source, and so have most commonly been used in combination with measured photometry and kinematics to estimate distances \citep[for example,][]{tf}. However,  recent large samples of UDGs \citep{vdk,munoz,koda} are confined to galaxies in galaxy clusters, which means that the distances are known. 

For galaxies with known or estimated distances we can use scaling relations to solve for the internal kinematics. This approach, utilising the Tully-Fisher relation, has been used for high surface brightness galaxies \citep{cole,gonzalez}. However,  the Tully-Fisher relation does not apply to UDGs because their morphology suggests that they are not rotating disk galaxies. Below we describe the application of a scaling relation that does apply to low velocity dispersion, pressure supported systems. We will demonstrate that using this approach we recover the measured velocity dispersions of the two UDGs that have been spectroscopically measured so far. We then apply the method to two large, published samples of UDGs and conclude that the total masses of these systems range from those of dwarf galaxies to Milky-Way like systems. We conclude that UDGs, as a class, cannot be thought of as either all dwarfs or all failed $L_*$ galaxies.

\section{Defining, Applying, and Testing the Method}

A series of studies have presented an extension of the Fundamental Plane relation \citep{djorg,d87}, which is referred to as the Fundamental Manifold (FM) in acknowledgment of its antecedent. The full range of known stellar systems, all luminosities, morphologies, and dark matter fractions, can be placed onto a single scaling relation \citep{z06a,z06b, z08}. A key feature of that relation is that the mass-to-light ratio, $\Upsilon_e$, within the half light or effective radius, $r_e$, is expressed as a polynomial that is a function of the internal kinematics, V, and the mean surface brightness within $r_e$, $I_e$,
\begin{equation}
\begin{split}
\log \Upsilon_e  &=   0.24\ ({\log {\rm V}})^2 + 0.12\ (\log I_e)^2  - 0.32\ {\log {\rm V}} \\
& \hspace{50pt} - 0.83\ {\log I_e} - 0.02\ \log {\rm V}I_e + 1.49,  
\end{split}
\label{eq:m2l}
\end{equation}
where the coefficients were empirically determined to minimise the scatter for a calibration sample and the kinematic term, 
V, is defined to be the combination of the line of sight velocity dispersion, $\sigma_v$, and the inclination corrected rotation speed, v$_{r}$, V$\equiv \sqrt{\sigma_v^2 + {\rm v}_{r}^2/2}$. 
 We are adopting the coefficients presented by \cite{z08}, but these need to be continually updated as more and better data become available. $\Upsilon_e$ is calibrated to the V-band, but the expression is accurate in any band to the degree that the colours of the galaxy do not deviate from the Solar colours, otherwise colour corrections are necessary. Given $\Upsilon_e$,  the Virial theorem and standard assumptions lead to a simple expression 
\begin{equation}
\log r_e = 2\log {\rm V} - \log I_e - \log \Upsilon_e - 0.75,
\label{eq:fm}
\end{equation}
where the constant 0.75 is empirically determined, again using a calibration sample \citep{zrev}.

\begin{figure}
\includegraphics[width=\columnwidth]{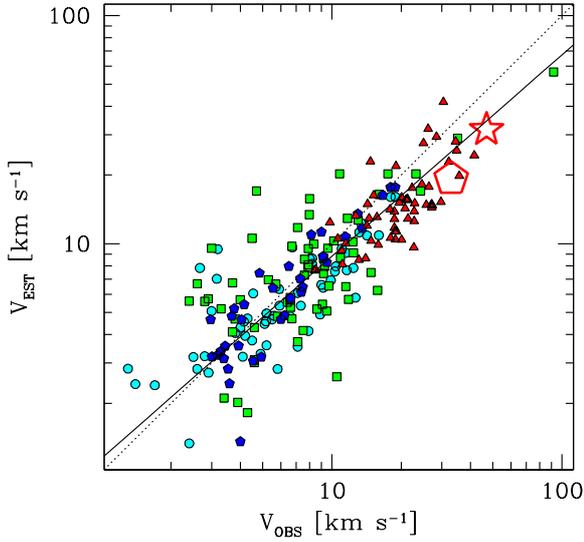}
\caption{Comparison of estimated kinematic terms, $V_{\rm EST}$, obtained by solving Equations \ref{eq:m2l} and \ref{eq:fm} to spectroscopically measured ones, V$_{\rm OBS}$. For almost all of the galaxies, those without measured rotation, the kinematic term is equivalent to the line of sight velocity dispersion, $\sigma_v$. A variety of samples are shown. The cyan circles represent the globular cluster sample of \citet{mclaughlin}, the dark blue pentagons the stellar cluster samples of 
\citet{z12,z13,z14},  the green squares the Local Group galaxies of \citet{mcc}, and the red triangles the ultracompact galaxies from \citet{mieske}. The large open symbols represent the two UDGs with measured dispersions (VCC 1287, from \citet{beasleya,beaslyb}, is represented by the pentagon and DF44 from \citet{vdk16}, by the star}). The dotted line is the 1:1 line and the solid line is a bisector fit \citep{isobe} to the data and is used to apply the small corrections that convert the estimate velocity dispersion to $\sigma_v^*$.
\label{fig:comp}
\end{figure}

With known $r_e$ and $I_e$, we use Equations \ref{eq:m2l} and \ref{eq:fm} to solve for $\Upsilon_e$ and V. 
The result of using this method on several sets of galaxies for which measured velocity dispersions exist for comparison are presented in Figure \ref{fig:comp}. We include faint Local Group galaxies from \cite{mcc}\footnote{The database is updated and made publicly available 
at http://www.astro.uvic.ca/$\sim$alan/Nearby\_Dwarf\_Database.html }, globular clusters from the compilation by \cite{mclaughlin}, a smaller set of stellar clusters with more precise velocity dispersions \citep{z12,z13,z14}, and compact dwarf galaxies from \cite{mieske}. For a few of the galaxies (in the Local Group sample)  there is measured rotation and we include that in V as defined above. For all other systems V$ = \sigma_v$. The comparison yields a mean value of the kinematic term that is on average only 2\% lower than the measured one, but on a case by case basis has an uncertainty of $\sim$50\%. Because slight systematic deviation from the expectation is visible, we fit a correction term as shown in Figure \ref{fig:comp} using the bisector fit from \cite{isobe} and only the data described above (not the UDG measurements discussed next). The corrected velocity dispersion, $\log \sigma_v^* = (\log \sigma_v - 0.061)/0.883$, is what we use. The existence of a systematic correction shows that the FM calibration can be improved.

\begin{figure}
\includegraphics[width=\columnwidth]{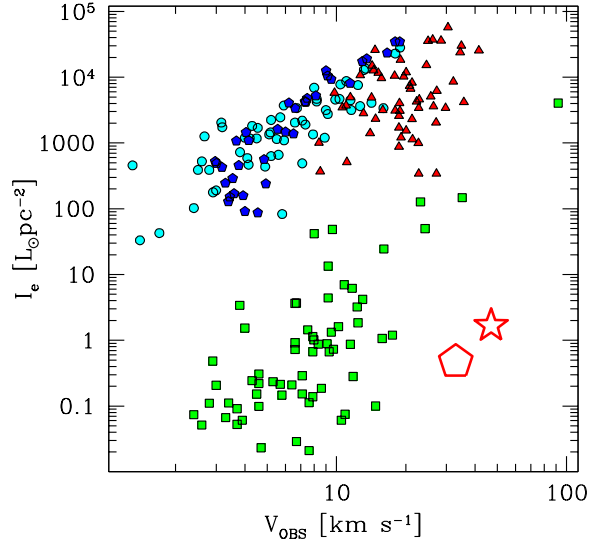}
\caption{How the $V-I_e$ space is sampled by the stellar systems in Figure \ref{fig:comp}. Symbols as in Figure \ref{fig:comp}. The UDGs lie in a region of the space that has not previously been sampled. Therefore, the use of the empirical calibration for $\Upsilon_e$ is a slight extrapolation when applied to these galaxies. Confirming the adopted behaviour in this region of the space with spectroscopic measurements of more UDGs would solidify the argument for the applicability of the FM.}
\label{fig:sb}
\end{figure}

We show in Figure \ref{fig:sb} why the current application needs to be treated with the caution appropriate for any extrapolation. In the Figure we plot the same objects as in Figure \ref{fig:comp}, but this time in the $V-I_e$ space. This is the space over which Equation \ref{eq:m2l} is calibrated. The UDGs lie in an area where there has been no previous constraint. However, considering either variable alone there are analogs in the calibration sample and so, unless the relation between $\Upsilon_e$ and the parameters is of much higher order in the UDG portion of this space than over the calibrated region, the extrapolation is minor relative to the full range covered by the parameters. Even so, spectroscopic confirmation for a set of UDGs is essential. We have two systems so far that can be used for such a test.

We  evaluate $\sigma_v^*$ for the two UDGs with published velocity dispersions \citep[VCC 1287, from six globular cluster velocities, and DF 44, from integrated light;][respectively]{beasleya,vdk16}. For VCC 1287 we obtain a velocity dispersion of 24 km s$^{-1}$ in comparison to the published value of 33$^{+16}_{-10}$ km s$^{-1}$, while for DF44 we obtain 43 km s$^{-1}$ in comparison to the published value of 47$^{+8}_{-6}$ km s$^{-1}$. Our estimates are within 1$\sigma$ of the published results.

\begin{figure}
\includegraphics[width=\columnwidth]{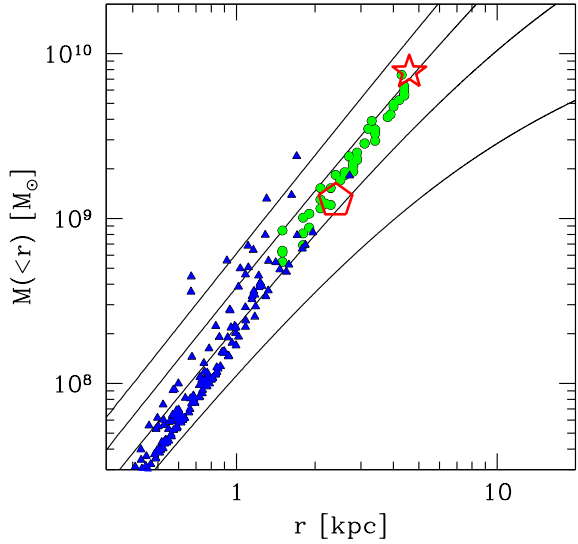}
\caption{Comparison of estimates of the enclosed mass within $r_e$ for the \citet{vdk}
sample of UDGs in the Coma cluster (green circles and red star) and the \citet{munoz} sample in the Fornax cluster (blue triangles), to NFW
enclosed mass profiles for four different values of $M_{200}, 10^{10}, 10^{11}, 10^{12},$ and $10^{13} M_\odot$. The red star represents DF44, the only object in these two samples with a spectroscopically measured velocity dispersion. The Coma UDGs span masses of $10^{11}$ to $10^{12} M_\odot$, while the Fornax ones span mostly
$10^{11}$ to $10^{10} M_\odot$. The red pentagon is VCC 1287 and is included for completeness. The trend is for 
the physically larger systems to also be the more massive. We discuss the outliers in the Fornax sample in the text.}
\label{fig:mass}
\end{figure}

Having established a basis for the use of the FM to estimate UDG velocity dispersions, we proceed to examine larger samples.
For the set of Coma UDGs \citep{vdk}, we estimate velocity dispersions ranging from 20 to 43 km s$^{-1}$. DF44 is estimated to have the second largest velocity dispersion in the sample. Regarding $\Upsilon_e$, we obtain estimates ranging from 11 to 100 in solar units. These are all dark matter dominated systems.
We compare the distributions of masses enclosed within $r_e$, $M(< r_e)$, using $M(<r_e) = 9.3\times10^{5} \sigma_v^2 r_e$ \citep{wolf} to enclosed mass curves for NFW mass profiles \citep{nfw} in Figure \ref{fig:mass}. This comparison requires some care because the concentration of a halo of a given mass is dependent on the adopted cosmological parameters. We have utilised the formulae provided by \cite{ludlow} appropriate for a Planck cosmology. The tight sequence of points arises because we have assumed that $V$ is a function of $r_e$ and $I_e$ and because these galaxies were selected to be in a narrow range of surface brightness. Therefore, a sequence of $r_e$ values maps nearly directly onto a sequence of $V$, and the enclosed mass is simply a function of $r_e$ and $V$. If UDGs do all lie on the FM, then $r_e$ maps fairly closely to total mass.

For the set of Fornax UDGs \citep{munoz}, we also evaluate enclosed masses and present results in Figure \ref{fig:mass}. These galaxies are mostly physically smaller than the Coma UDGs and span the enclosed mass profiles of halos with $M_{200}$ $\sim 10^{10}$ to $10^{11} M_\odot$. The Fornax sample has some outliers that appear to be non-physical, such as those with masses $> 10^{13} M_\odot$. There are various potential explanations for these objects. First, these may be objects for which the FM is not applicable. They may occur because they truly are FM outliers or because they are in a dynamical state that violates the Virial assumptions, for example they be tidally disrupting. Second, they may not be in the Fornax cluster. This would invalidate the assumed distance and  our estimates of $\sigma^*_v$ and $\Upsilon_e$. Third, there may be errors in the measured half light radii or total magnitudes. Regardless of which of these scenarios is correct, following these objects up is a priority. The three Fornax UDGs with the most extreme properties ($M(<r) > 10^9 M_\odot$ and $r_e < 2$ kpc) are NGFS033429-353241, NGFS033807-352624, and
NGFS033913-352217. 

Finally, we have also included the Virgo galaxy, VCC 1287, for comparison in Figure \ref{fig:mass}. Our estimate for the enclosed mass within $r_e$ for this galaxy is $1.3 \times 10^9$ M$_\odot$, lower than the published estimate \citep[$2.6^{+3.4}_{-1.3} \times 10^9$ M$_\odot$;][]{beasleya} but just within 1$\sigma$ of that value.

\section{Conclusions}

We place ultra diffuse galaxies (UDGs) on an existing galaxy scaling relation to estimate their line of sight velocity dispersions. We find that the two UDGs with spectroscopically measured dispersions satisfy the scaling relation sufficiently well that our estimated velocity dispersions are within the spectroscopic $1\sigma$ uncertainties. Assuming that these two galaxies signify that UDGs as a class satisfy the scaling relation, we derive velocity dispersions, and the related enclosed dynamical masses, for the full set of UDGs in the Coma \citep{vdk} and Fornax \citep{munoz} clusters. We reach the following conclusions:

$\bullet$ DF44 appears to lie in a massive ($\sim 10^{12} M_\odot$) as found by \cite{vdk16}, but is not representative of UDGs. It lies at the upper end in size and enclosed mass of the known UDGs. Therefore, while it appears to be what has been referred to as a failed L$_*$ galaxy, it is not typical of UDGs. 

$\bullet$ Different samples of UDGs can be quite different. The Coma and Fornax samples have little overlap in half light radii and that translates to little overlap in enclosed mass and total mass. The Fornax UDGs generally have $M_{200} < 10^{11} M_\odot$ and so can plausibly be referred to as dwarf galaxies (e.g. LMC-like and smaller). However, they are not fully representative of UDGs because the Fornax sample does not include an object as extreme as DF44. 

We conclude that UDGs are neither all dwarfs nor all failed L$_*$ galaxies. Subsequent spectroscopic measurements of velocity dispersions will help resolve whether the Fundamental Manifold can be used to reliably estimate velocity dispersions of UDGs that have known distances. If so, we will then be able to explore a number of questions regarding the nature of these objects without the overwhelming burden of obtaining spectroscopic velocity dispersions for large samples of these challenging galaxies.

\section{Acknowledgements}

DZ acknowledges financial support from NSF AST-1311326 and the University of Arizona.

%%%%%%%%%%%%%%%%%%%%%%%%%%%%%%%%%%%%%%%%%%%%%%%%%%



%\bibliographystyle{mnras}


%%%%%%%%%%%%%%%%%%%%%%%%%%%%%%%%%%%%%%%%%%%%%%%%%%

% Don't change these lines
\bsp	% typesetting comment
\label{lastpage}
\end{document}